\documentstyle[preprint,floats,epsf,eqsecnum,aps,tighten]{revtex}
\begin{document}
\draft

\preprint{
 \parbox{1.5in}{\leftline{JLAB-THY-97-28}
                \leftline{WM-97-110}
                \leftline{nucl-th/97????} }  }

\title{Lagrangian with off-shell vertices and field redefinitions.  }

\author{
 J.~Adam, Jr.$^{1,2}$, Franz~Gross$^{1,3}$ and  J.~W.~Van
Orden$^{1,4}$ }
\address{
$^1$Jefferson Lab,
12000 Jefferson Avenue, Newport News, VA 23606\\
$^2$ Nuclear Physics Institute, Czech Academy of Sciences,
CZ-25068 {\v R}e{\v z} near Prague, Czech Republic \\
$^3$ Department of Physics,
College of William and Mary, Williamsburg, VA 23185 \\
$^4$Department of Physics, Old Dominion University, Norfolk,
VA 23529}

\date{\today}
\maketitle

\begin{abstract}
Meson exchange diagrams following from a lagrangian
with off-shell meson-nucleon couplings are
compared with those generated from conventional
dynamics. The off-shell interactions can be transformed
away with the help of a nucleon field redefinition.
Contributions to to the $NN$- and $3N$-potentials and nonminimal
contact e.m.\ meson-exchange currents are discussed,
mostly for an important case of scalar meson exchange.
\end{abstract}



\section{Introduction}

A nucleon-nucleon interaction is efficiently parametrized
in terms of meson exchange diagrams. Apart from the largely model
independent one pion exchange contribution, a small family
of heavier mesons ($\omega$, $\rho$, $\sigma$, sometimes also
$\delta$ and $\eta$) effectively summing and representing the
multipole pion exchanges and excitations of nucleon resonances
in intermediate states is considered. The structure of the couplings
of these (effective) mesons to nucleons is not very well known
and is treated phenomenologically:
vertices are usually taken in some simple form, neglecting effects
due to possible transition of the nucleons off their mass shell,
 and coupling constants and coupling parameters are fitted to data.

Some relativistic calculations \cite{StG,Zim} have found that
including  off-shell extensions of these meson-nucleon vertices might
give a more effective description. In particular, Stadler and Gross
\cite{StG} have shown in a covariant spectator formalism that using
a  scalar-nucleon-nucleon ($sNN$) coupling with off-shell extension
can give a
reasonable triton binding energy (without explicit three-nucleon
forces), and at the same time improve the fit to $NN$ data as compared
to that of similar model without off-shell coupling. Zimanyi and
Moszkowski have shown that a  lagrangian with a derivative
(off-shell)  $sNN$ coupling
improves the description of nuclear matter and finite nuclei in the
relativistic mean-field approximation.  And a number of authors have
studied off-shell couplings using sidewise dispersion relations, which
suggest that the off-shell behavior should be related to $\pi N$
scattering and higher nucleon resonances\cite{DP,SW}.

Here, we would like to demonstrate how we can use nucleon field redefinition
to translate the dynamical model of \cite{StG} into
 a model with nonlinear couplings with standard on-shell vertices.
First, as an example, we demonstrate the nontrivial dynamical content of
off-shell vertices using the well known $\sigma$-model. Then, in
Sec.~III,  we consider
the scalar-exchange part of the Stadler and Gross  \cite{StG} model.
We show that the off-shell coupling can be removed via a redefinition of
the nucleon field and identify the nonstandard nonlinear
strong and electromagnetic (e.m.) vertices. In leading order
the difference between the model with off-shell coupling and the
standard one is represented by triangle  and bubble diagrams
for the $NN$-interaction, scalar-scalar
exchange three-nucleon potential, and a contact (seagull) meson exchange
current. In Sec.~IV we give explicit expressions for these contributions
and argue that they should be estimated numerically. Finally, in Secs.~V
and VI we discuss electromagnetic interactions and some features of the
nonrelativistic limits of these interactions.  In the appendix we show that
similar considerations also apply for pseudoscalar and vector
exchanges.

\section{$\sigma$-model as simple example}

To illustrate the rich dynamical content of off-shell couplings, let
us consider the simple example of a $\sigma$-model. The standard lagrangian
of the linear $\sigma$- model is
\begin{eqnarray}
&&{\cal L}  =  {\cal L}_{N,0}^{kin} (\psi ) + {\cal L}^{kin} (\sigma,\pi)
- g \bar{\psi} \Phi \psi - \frac{1}{2} m_{\Phi}^2\, \Phi^2 +
 V(\Phi^2) \, ,
\label{llinsig}\\
&&{\cal L}_{N,0}^{kin} (\psi ) =
 \frac{i}{2} \bar{\psi} \gamma^{\mu} ( \partial_{\mu} \psi) -
 \frac{i}{2} ( \partial_{\mu}  \bar{\psi}) \gamma^{\mu}  \psi
   \, , \label{lnkin0}\\
&&{\cal L}^{kin} (\sigma,\pi ) =
  \frac{1}{2} \partial_{\mu} \sigma \partial^{\mu} \sigma +
  \frac{1}{2} \partial_{\mu} \vec{\pi} \partial^{\mu} \vec{\pi}
   \, ,
\label{lsigpikin}
\end{eqnarray}
where $\Phi = \sigma + i \gamma^5 \vec{\tau} \cdot \vec{\pi}$.
The nonlinear $\sigma$-model results from replacing $\Phi$ by a
nonlinear function of the pion field with  constant norm
$\Phi^2= \Phi \Phi^* = f_{\pi}^2$. The common choice is
\begin{equation}
 \Phi(\varphi) = f_{\pi} \exp (2 i \gamma^5 \varphi)  \, ,
\label{fi1}
\end{equation}
with $ \vec{\varphi} = \vec{\pi}/(2 f_{\pi})$ and
$\varphi= \vec{\tau} \cdot \vec{\varphi}$, which leads to a lagrangian
\begin{equation}
{\cal L}  = {\cal L}_{N,0}^{kin} (\psi )
 + \frac{1}{4} Tr \, \partial_{\mu} \Phi \partial^{\mu} \Phi^*
- f_{\pi} g \bar{\psi} \exp (2 i \gamma^5 \varphi)  \psi   \, ,
\label{lnonlin1}
\end{equation}
where the trace is taken in flavor space. In the lowest
order in pion fields the second term  gives the kinetic energy
for massless pions, while higher orders generate complicated nonlinear
pion self-interactions. The last term of (\ref{lnonlin1}) in the lowest
order generates the nucleon mass $m= g f_{\pi}$, the first order
in the pion field is the pseudoscalar  $\pi NN$ coupling, and  higher
orders represent multipion contact terms.

Alternatively, $\Phi(\varphi)$ can be taken in the form
\begin{equation}
 \Phi(\varphi) = f_{\pi}
   \frac{1- \vec{\varphi}^{\, 2} + 2 i \gamma^5 \varphi}
   {1+ \vec{\varphi}^{\, 2}} = f_{\pi} + 2 f_{\pi}
 \frac{- \vec{\varphi}^{\, 2} +
 i \gamma^5 \varphi}{1+ \vec{\varphi}^{\, 2}} \,  ,
\end{equation}
which yields
%
\begin{equation}
 {\cal L}  =  {\cal L}_{N,0}^{kin} (\psi ) +
\frac{1}{2 (1+ \vec{\varphi}^{\, 2})^2}
 \partial_{\mu} \vec{\pi} \, \partial^{\mu} \vec{\pi}
- f_{\pi} g \bar{\psi} \psi +
  \frac{2 f_{\pi} g }{(1+ \vec{\varphi}^{\, 2})}  \bar{\psi}
 (\vec{\varphi}^{\, 2} - i \gamma^5 \vec{\tau} \cdot
\vec{\varphi}\,  )\psi \,  ,
\label{nonlin2}
\end{equation}
We can eliminate the higher order nonlinear terms from the $\pi NN$
couplings by redefining the nucleon field as follows:
\begin{equation}
  \psi(x)= \sqrt{1+ \vec{\varphi}^{\, 2} (x) } \, \psi'  (x) \, ,
\label{psi2}
\end{equation}
which gives the equivalent lagrangian
\begin{eqnarray}
 {\cal L} & = & {\cal L}_{N,0}^{kin} (\psi' )  - f_{\pi} g
\bar{\psi'} \psi' +
\frac{1}{2 (1+ \vec{\varphi}^{\, 2})^2}
 \partial_{\mu} \vec{\pi} \, \partial^{\mu} \vec{\pi}
 +  2 f_{\pi} g \bar{\psi'}
 (\vec{\varphi}^{\, 2} - i \gamma^5 \vec{\tau} \cdot
\vec{\varphi} ) \psi'
 \nonumber\\
&&  + \frac{1}{2} \bar{\psi'}\, \vec{\varphi}^{\, 2}
  ( i \gamma^{\mu} \partial_{\mu} \psi' - m \psi') -
 \frac{1}{2} ( i (\partial_{\mu} \bar{\psi'}) + m \bar{\psi'})
  \vec{\varphi}^{\, 2} \,  \psi'  \,  ,
\label{nonlin3}
\end{eqnarray}
Note that {\it all of the $\pi NN$ interaction terms involving the
coupling of three or more pions to a nucleon in the original lagrangian
(\ref{nonlin2}) have been replaced by terms which are only linear or
quadratic in the pion field\/}, but which involve couplings to off-shell
nucleons.

We now emphasize two points.  The two lagrangians (\ref{nonlin2}) and
(\ref{nonlin3}) are equivalent {\it only\/} if {\it all\/} of the
nonlinear terms in the factor
$f=1/(1+\vec{\varphi^{\,2}})$ are retained.  If this
factor is truncated to lowest order,
$f\simeq 1 - \vec{\varphi^{\,2}}$,
the lagrangians are {\it no longer equivalent\/}.  The
second point is that the lagrangian (\ref{nonlin3}) is deceptively
simple. In particular, at most two pions couple to a nucleon at any point.
Still, it contains all the complexity of the nonlinear $\sigma$-model with
all of its multipion contact terms. Naively, one might argue that the
off-shell term is unimportant in realistic nuclear applications,
since it is nonzero only for off-shell nucleons. Indeed, in the
lowest semiclassical order of two meson exchange it does not contribute.
But as momentum loops are included the off-shell vertices give
results compatible with original nonlinear $\sigma$-model.
Iterating the off-shell vertices of
(\ref{nonlin3}) along the same nucleon line immediately generates
multipion contact terms with any number of pions,
since the off-shell factors ($\hat{p} -m $ in momentum space) of the vertex
cancel the  attached nucleon propagators.

However, it
might be nontrivial to define the effective theory based on the
lagrangian (\ref{nonlin3}). This is because the nucleon field has 
complicated transformation properties and one has to be careful 
in defining the regularization
so that the symmetry is preserved and the theory is equivalent to the
usual nonlinear $\sigma$-model. We are interested in a more
phenomenological approach in which only a limited set of Feynman
diagrams (regularized by ad hoc hadronic form factors) is used
in a kernel of a dynamical equation. In this case, the dynamics defined
by (\ref{nonlin3}) is no longer fully equivalent to the $\sigma$-model,
but it is still clearly distinct from the standard prescriptions
employing the linearized form of the $\pi NN$ interaction.  

We now turn to a discussion of off-shell couplings associated with the
exchange of scalar particles.

\section{Field redefinitions for off-shell scalar coupling}

Let us now consider the scalar coupling with the off-shell extension
which plays an important role in the dynamical model discussed in
Ref.~\cite{StG}. We will first show how the off-shell part of the $sNN$
coupling can be removed with the help of the nucleon field redefinition.
The field redefinition generates the nonlinear scalar-nucleon and
photon-scalar-nucleon vertices. In the next sections, we present
the contributions of these vertices to the $NN$ potential, the $3N$
potential and the e.m.\ exchange currents in the leading order
beyond conventional results.

The part of lagrangian relevant to our discussion is
\begin{eqnarray}
  &&{\cal L} = {\cal L}^{kin}_B + {\cal L}_{\gamma ss}
  + {\cal L}^{kin}_N (\psi ) + {\cal L}_{sNN} (\psi )
 + {\cal L}_{\gamma NN} (\psi ) + {\cal L}_{\gamma NNs}(\psi )    \, ,
 \label{lagrs0}\\
 &&{\cal L}^{kin}_N (\psi ) =
 \frac{i}{2} \bar{\psi} \gamma^{\mu} ( \partial_{\mu} \psi) -
 \frac{i}{2} ( \partial_{\mu}  \bar{\psi}) \gamma^{\mu}  \psi
   - m \bar{\psi} \psi \, , \label{lnkin}\\
  &&{\cal L}_{sNN} (\psi ) =   g_s \bar{\psi} \Phi_s \psi
 +  \frac{a_s}{2}  \bar{\psi}   \Phi_s
  ( i \gamma_{\mu} \partial^{\mu} \psi - m \psi  ) -
  \frac{a_s}{2}
  ( i (\partial^{\mu} \bar{\psi}) \gamma_{\mu} + m \bar{\psi}  )
   \Phi_s \psi       \, ,
  \label{lsnn}\\
  &&{\cal L}_{\gamma NN} (\psi )  =
    \bar{\psi} \Lambda^{\mu}  \psi \, A_{\mu} \ \ , \,
   \Lambda^{\mu} =  \Lambda_0^{\mu}+ \Delta  \Lambda^{\mu} \,
\label{lgnn} \\
  &&{\cal L}_{\gamma NNs} (\psi )  =  \frac{a_s}{2} \bar{\psi}
   \left\{ \Lambda_0^{\mu} \, ,
      \Phi_s \right\} \psi \, A_{\mu} \, \, ,
\label{lgnns}
\end{eqnarray}
where ${\cal L}^{kin}_B $ includes all of the kinetic terms for the bosons
(scalars and photons), ${\cal L}^{kin}_N (\psi )$ is the nucleon
kinetic lagrangian with the mass term, and ${\cal L}_{\gamma ss}$
is a photon-scalar vertex.

The function $\Phi_s$ contains an
isospin matrix for the mesons with nonzero isospin (i.e., $\Phi_s =
\vec{\tau} \cdot \vec{\Phi}_s $). The parameter $a_s$ in the off-shell
part of the $sNN$ vertex is related to the parameter $\nu_s$ of \cite{StG}
through $a_s= \nu_s g_s /m$ and the $sNN$ vertex function in the momentum
space reads
\begin{equation}
 \tilde{\Gamma}_s (p', p) =  1 + \frac{\nu_s}{2m} \left( \hat{p}' +
   \hat{p} - 2m \right) \, ,
\label{offps}
\end{equation}
In (\ref{lgnn}) we have separated a minimal part of the $\gamma NN$ vertex
\begin{equation}
\Lambda_0^{\mu} = \frac{e}{2} (1 + \tau^3)\, \gamma^{\mu}  \, ,
\label{vergnn0}
\end{equation}
with the proton charge $e > 0$,
from the remaining purely transverse one $\Delta  \Lambda^{\mu}$.
The minimal contact (seagull) interaction is contained in
${\cal L}_{\gamma NNs}(\psi )$. It is obtained by minimal substitution
of the derivatives of the nucleon field in (\ref{lsnn})
\begin{eqnarray}
 i \gamma^\mu \partial_\mu \psi & \to & \Lambda_0^{\mu}\psi \, , \\
 - (i \partial_\mu \bar{\psi}\gamma^\mu ) & \to &
 \bar{\psi} \Lambda_0^{\mu} \, .
\end{eqnarray}
The nonminimal part of the photon-nucleon coupling $\Delta  \Lambda^{\mu}$
has been given in a framework of effective meson-nucleon theories by
Gross and Riska \cite{GR}. In simplified models, $\Lambda_0^{\mu}$
and  $\Delta  \Lambda^{\mu}$ are often taken as the Dirac  and
Pauli parts of the $\gamma NN$ vertex, proportional
to  $F_1$ and $F_2$,  respectively.

The similarity of (\ref{lnkin}) and (\ref{lsnn}) suggests that
the off-shell vertex can be removed by the nucleon field redefinition
\cite{Zim}
\begin{eqnarray}
 \psi(x) &=& F_s(\Phi_s (x)) \, \psi'(x) \, , \\
 \bar{\psi}(x) &=&  \bar{\psi}'(x) F_s(\Phi_s (x))   , \\
   F_s (x)& = &1/\sqrt{1+ a_s \Phi_s(x) } \, .
\label{fscal}
\end{eqnarray}
In terms of the new field $\psi'$ the lagrangian (\ref{lagrs0}) reads
\begin{eqnarray}
  {\cal L} & = & {\cal L}^{kin}_s + {\cal L}^{kin}_N (\psi' )+
  {\cal L}_{\gamma ss} \nonumber\\
&& + g_s \bar{\psi'} \frac{\Phi_s}{1+ a_s \Phi_s}\psi'
 +  \frac{i}{2} \bar{\psi'} \gamma^{\mu}
 \left(    F_s^{-1}  (\partial_{\mu} F_s ) -
   (\partial_{\mu} F_s )  F_s^{-1}     \right) \, \psi' \nonumber\\
&&
 + \bar{\psi}' F_s \, \Lambda^{\mu} \, F_s \psi' \, A_{\mu}
   + \frac{a_s}{2} \bar{\psi}'  F_s \,
   \left\{ \Lambda_0^{\mu} \, , \Phi_s \right\} \,
    F_s   \psi' \, A_{\mu} \, \, .
\label{lagrs}
\end{eqnarray}
As in the previous section, the off-shell meson-nucleon vertex can be
transformed away in favor of complicated nonlinear contact
interactions. The leading term of the nonlinear scalar-nucleon term
is the conventional scalar-nucleon vertex, the terms with derivatives
of the scalar field $(\partial_{\mu} F_s )$ contribute only for the scalar 
with nonzero isospin.

While the off-shell interaction can be easily
included in covariant dynamical equations for $NN$ and $3N$ systems,
one cannot take the nonlinear vertices to all orders, and hence in
practice the two forms of the lagrangian are not equivalent (although
they are in principle).  In an energy region where the effective
meson-nucleon description is valid, the importance of the  multimeson
exchanges and/or exchanges of heavy mesons  decreases with increasing
summed mass of exchanged mesons. In particular, since the mass of the
effective scalar meson is typically $m_s \approx 500$ MeV, it might be
interesting to consider the effects of nonlinearities up to second order
in the scalar field, and see if the differences between a truncated
version of (\ref{lagrs}) and the orignial (\ref{lagrs0}) can be
explained mostly by the second order terms.

Since the original lagrangian (\ref{lagrs0}) depends on the off-shell
parameter $\nu$ (or $a$), so also will  physical observables,
like phase shifts and the $3N$ binding energy (parameters of the model
\cite{StG} are fine tuned to leave the deuteron binding energy
unchanged). However, the leading order one meson exchange interaction
does not depend on $\nu$, at least for external legs on-shell.
But nucleon interactions with
two exchanged mesons, or interactions with external fields with at
least one meson simultaneously exchanged, are already dependent on
the choice of $\nu$. We present the corresponding $\nu$-dependent operators
below, first for the two-scalar exchange $NN$ and $3N$ potentials and then
for the one-scalar exchange e.m.\ current.

\section{Two-scalar exchange and nuclear interaction}

Since the local field redefinition does not change
the physical $S$-matrix elements, the $NN$-interaction generated by the
lagrangian (\ref{lagrs}) with nonlinear $sNN$ couplings should give the
same result as the lagrangian (\ref{lagrs0}) where the nonlinear
couplings are replaced by off-shell couplings. It is easy to see that the
leading, second order contributions due to a single scalar exchange are the
same in both frameworks.  In this section we show that this is also true of
the fourth order, two scalar exchange terms.  The demonstration
illustrates explicitly how the off-shell couplings generate higher order
contact terms.

\begin{figure}[t]
\begin{center}
\mbox{
   \epsfysize=1.3in
\epsfbox{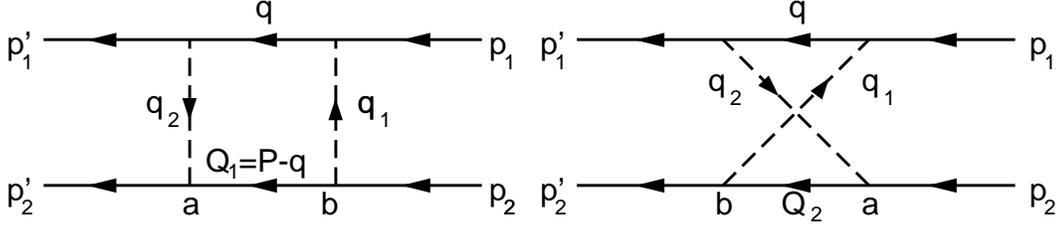}
}
\end{center}
\caption{Box and crossed box contributions to the two-meson
         exchange $NN$ potential.}
\label{boxcros}
\end{figure}

Two scalar exchange contributions with off-shell couplings are represented
by the box and crossed box contributions of Fig.~\ref{boxcros}.
%
Using (\ref{offps}) the corresponding amplitudes
are
\begin{eqnarray}
{\cal M}_{box} = i g_s^4 \int \frac{d^4 q}{(2 \pi)^4}&&
 D(q_1) D(q_2) \,
  \bar{u}(p'_1) \tilde{\Gamma}_s^a (p'_1,q) G(q)
  \tilde{\Gamma}_s^b (q,p_1) u(p_1) \nonumber\\
 &&\times  \bar{u}(p'_2) \tilde{\Gamma}_s^a (p'_2,Q_1) G(Q_1)
 \tilde{\Gamma}_s^b (Q_1,p_2) u(p_2)
 \, ,
\label{mbox}\\
{\cal M}_{cross} = i g_s^4 \int \frac{d^4 q}{(2 \pi)^4}&&
 D(q_1) D(q_2) \,
  \bar{u}(p'_1) \tilde{\Gamma}_s^a (p'_1,q) G(q)
  \tilde{\Gamma}_s^b (q,p_1)  u(p_1) \nonumber\\
 && \times \bar{u}(p'_2)  \tilde{\Gamma}_s^b (p'_2,Q_2) G(Q_2)
 \tilde{\Gamma}_s^a (Q_2,p_2)  u(p_2)
 \, ,
\label{mcross}
\end{eqnarray}
where $G(p)= 1/(m -\hat{p} - i \epsilon)$ is the nucleon
propagator, $D(q)= 1/(m_s^2 - q^2 -  i \epsilon)$ is the scalar
propagator, and the momenta are defined in Fig. \ref{boxcros}.

\begin{figure}[t]
\begin{center}
\mbox{
   \epsfysize=1.0in
\epsfbox{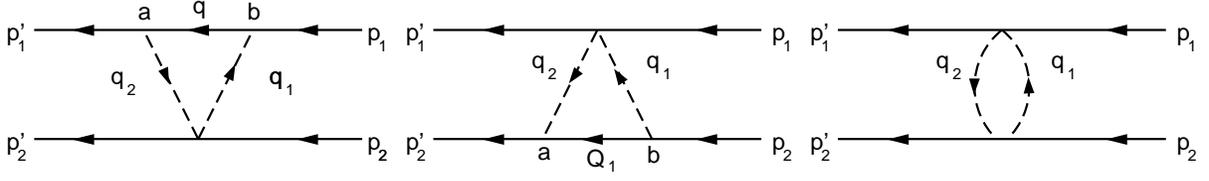}
}
\end{center}
\caption{Triangle and bubble contributions to the two-scalar
         exchange $NN$ potential generated by the contact
         interactions in the transformed lagrangian.}
\label{tribub}
\end{figure}

The sum of these amplitudes should be equal to the sum of
the box and crossed-box diagrams with the on-shell vertex
$\Gamma_s=1$ and the triangle and bubble diagrams of Fig.~\ref{tribub}.
These triangle and bubble diagrams are generated from the
quadratic contact vertex
\begin{equation}
{\cal L}_{ssNN} (\psi') =
 - \frac{g_s^2 \nu_s}{m}  \bar{\psi'} \Phi_s^2 \psi'
  +  \frac{g_s^2 \nu_s^2}{4 m^2}
 \bar{\psi'}\gamma^{\mu} \vec{\tau} \cdot \vec{\Phi}_s \times
 \partial_{\mu}  \vec{\Phi}_s  \psi'   \, ,
\label{lssnn}
\end{equation}
which follows from the Taylor decomposition of the nonlinear vertex
in (\ref{lagrs}). In the momentum representation the quadratic
contact vertex is
\begin{equation}
 \Gamma^{ab}(q_2,q_1) = - 2 g_s^2 \frac{\nu_s}{m}
 \left( \delta^{ab} + i \epsilon^{abc} \tau^c
 \frac{\nu_s}{8m}( \hat{q}_1+ \hat{q_2}) \right)\, ,
\label{scal2con}
\end{equation}
where $a$ and $b$ are isospin indices associated with the incoming
$\Phi^b(q_1)$ and outgoing
$\Phi^a(q_2)$ scalar fields.
%
The amplitudes corresponding to these diagrams in Fig.~\ref{tribub} are
\begin{eqnarray}
{\cal M}_{tri} = - i g_s^4 \int&& \frac{d^4 q}{(2 \pi)^4}
 D(q_1) D(q_2)\,  2 \frac{\nu_s}{m}
 \Biggl\{ \delta^{aa}
\left[ \bar{u}(p'_1) G(q) u(p_1) \bar{u}(p'_2) u(p_2) \right.\nonumber\\
&&+\left. \bar{u}(p'_1) u(p_1) \bar{u}(p'_2) G(Q_1) u(p_2) \right]
+ \frac{\nu_s}{2m}\left[ \bar{u}(p'_1) G(q) \tau^a u(p_1)
 \bar{u}(p'_2) \hat{Q} \tau^a u(p_2) \right.\nonumber\\
&&\qquad \qquad\left.-\bar{u}(p'_1) \hat{Q} \tau^a u(p_1)
   \bar{u}(p'_2) G(Q_1) \tau^a u(p_2) \right] \Biggr\} \, ,
\label{tri}\\
{\cal M}_{bub} = i g_s^4 \int &&\frac{d^4 q}{(2 \pi)^4}
 D(q_1) D(q_2) \, 2 \frac{\nu_s^2}{m^2}
\Biggl\{ \delta^{aa} \bar{u}(p'_1) u(p_1) \bar{u}(p'_2) u(p_2)\nonumber\\
 &&\qquad\qquad + \frac{\nu_s^2}{8m^2} \bar{u}(p'_1) \hat{Q} \tau^a u(p_1)
   \bar{u}(p'_2) \hat{Q} \tau^a u(p_2) \Biggr\} \, .
\label{bub}
\end{eqnarray}
where we have introduced $Q= (q_1 + q_2)/2$.
The equivalence of ${\cal M}_{box} + {\cal M}_{cross}$ to the sum
of conventional box and crossed-box diagrams and
${\cal M}_{tri} + {\cal M}_{bub}$ given above follows from simple
algebra. The substitution
$ q \to -q + p'_1 + p_1$ in the integral, which  leads to replacements
$q_1, q_2, Q_1, Q_2, Q \to -q_2, -q_1, Q_2, Q_1, -Q$,  is
useful in the proof.

If the scalar field has a zero isospin, $ \delta^{ab}, \delta^{aa} \to 1$
and terms with other isospin structures disappear.
The terms containing $\hat{Q}$ in Eqns.~(\ref{tri}) and (\ref{bub}) are
suppressed by extra powers of $1/m$. Hence,  at least in the leading
order, the
$\nu_s$-dependent two-scalar exchange contributions have very simple
structure. The $\nu_s$-dependent two-scalar exchange $NN$ potentials
follow from (\ref{tri}) and (\ref{bub}) by a straightforward nonrelativistic
reduction of the vertices, in particular $\bar{u}(p'\, ) u(p) \to 1$.

The triangle and bubble diagram contributions to the
$NN$ potential in the leading order in $1/m$ were recently constructed
by Rijken and Stoks \cite{RS} for pions and $\sigma$-mesons, where only
positive-energy nucleons were considered in intermediate states.
It would be interesting to check whether the contributions
derived above could account for a sizable part of dynamical
difference between a model with the standard $sNN$ coupling and a
model with its
off-shell extension. To this end one should  fix all
other meson parameters and fit the conventional model to the $NN$ data.
Then one could compare the effects of adding either an off-shell
vertex or the triangle and bubble interactions to the potential.

\begin{figure}[t]
\begin{center}
\mbox{
   \epsfysize=1.3in
\epsfbox{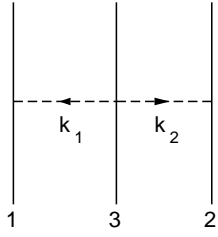}
}
\end{center}
\caption{Contribution to the trinucleon potential generated by the contact
         interactions in the transformed lagrangian. Two other diagrams with
         cyclic permutation of nucleons are not shown.}
\label{nnn}
\end{figure}

The two-scalar exchange also contributes to the three-nucleon potential
shown on Fig. \ref{nnn}.
%
 In particular, from the $\nu$-dependent quadratic
vertex (\ref{scal2con}) one gets
\begin{eqnarray}
 V_{3N} &=& 2 \frac{\nu_s}{m} \,  g_s^4  \, 
  D(k_1) D(k_2) \,  \bar{u}(p'_1) \tau^a u(p_1) \,
  \bar{u}(p'_2) \tau^b u(p_2) \nonumber\\
  &&  \bar{u}(p'_3) \left[ \delta^{ab} + i \epsilon^{cab} \tau^c
  \frac{\nu_s}{8m} (\hat{k_1} - \hat{k_2}) \right] u(p_3)
  + acycl \, .
\label{v3}
\end{eqnarray}
The potential simplifies when only lowest order in $v/c$
is retained, which means replacing all the vertices by unity.
This gives a very simple, central three-nucleon potential,
which is attractive for $\nu_s < 0 $, in agreement
with the increased binding  for negative  $\nu_s $ observed
in \cite{StG}.
Together with the triangle contributions to the $NN$ interaction
discussed above, this three-nucleon potential should account for part
of the large effect of the off-shell scalar coupling on the
triton binding energy \cite{StG}. It might be interesting to check this
numerically and to compare the importance of the
three-nucleon force to variations of the $NN$-interaction
due to  additional two-scalar exchanges.

\section{Electromagnetic interaction}

Since the e.m.\ part of the transformed lagrangian (\ref{lagrs})
also contains complicated nonlinear multimeson interactions,
the comparison of the e.m.\ observables calculated using nonlinear
models or models with off-shell couplings requires some care.
However, in the spirit of the previous section one can try to
explain a part of the difference by estimating the leading
order effects, which for the e.m.\ interaction are nonminimal
single-scalar exchange e.m.\ currents. Making the Taylor
decomposition of the e.m.\ part of (\ref{lagrs}) with the
help of $F_s \simeq 1 - a_s \Phi_s/2$ we get
\begin{eqnarray}
 {\cal L}_{e.m.} & = &
  \bar{\psi}' F_s \, \Lambda^{\mu} \, F_s \psi' \, A_{\mu}
   + \frac{a_s}{2} \bar{\psi}'  F_s \,
   \left\{ \Lambda_0^{\mu} \, , \Phi_s \right\} \,
    F_s   \psi' \, A_{\mu} \nonumber\\
 & \simeq &   \bar{\psi}' \Lambda^{\mu} \psi' \, A_{\mu}  -
   \frac{a_s}{2} \bar{\psi}'
   \left\{ \Delta \Lambda^{\mu} \, , \Phi_s \right\} \,
    \psi' \, A_{\mu} \, .
\label{lagrsem}
\end{eqnarray}
The minimal part of the contact $\gamma NNs$ vertex
proportional to $\Lambda_0^{\mu}$ disappears, since the
lagrangian (\ref{lagrs}) does not contain derivatives of the
scalar field in the linear $sNN$ vertex. Instead, a nonminimal,
purely transverse contact interaction appears. It is, of course,
not clear that the lagrangian with off-shell coupling from
which we start should have {\it a minimal} contact e.m.\ coupling.
But, in any case, the field definition (\ref{fscal}) transforms the nuclear
electromagnetic current (\ref{lgnn}) into an interaction
current, which to lowest order in the scalar field is
\begin{equation}
 \delta {\cal L}_{\gamma NNs} =  - \frac{a_s}{2}
 \bar{\psi}' \left\{ \Lambda^{\mu} \, , \Phi_s \right\} \,
    \psi' \, A_{\mu} \, .
\label{ldelgnns}
\end{equation}
This must be added to the transformation of whatever other contact 
interaction replaces (\ref{lgnns}) in the original lagrangian.
\begin{figure}[t]
\begin{center}
\mbox{
   \epsfysize=1.3in
\epsfbox{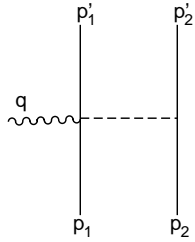}
}
\end{center}
\caption{Contribution to the electromagnetic meson-exchange
         current generated by the contact
         interactions in the transformed lagrangian. Only the
         diagram with the contact vertex attached to the first
         nucleon is shown.}
\label{mec}
\end{figure}

One can easily write down the interaction current $\delta
j^{\mu}_{s,con}(q)$, corresponding to (\ref{ldelgnns}) and given in
Fig.~\ref{mec}
\begin{eqnarray}
 \delta j^{\mu}_{s,con}(q) =&& \frac{g_s a_s}{2} \, D_s(p_2'- p_2)\,
 \bar{u} (p_1') \left( \tau^a \Lambda^{\mu}(p_1+q,p_1)
 + \Lambda^{\mu}(p_1',p_1'-q) \tau^a \right) u(p_1) \nonumber\\
&&\quad\times\bar{u} (p_2') \tau^a u(p_2) \, ,
\label{jcon}
\end{eqnarray}
where for a scalar meson with isospin $I=0$ one has to
replace $\tau^a \to 1$.  This current should account for
some part of the difference between the e.m.\ observables, e.g., deuteron
form factors, calculated in the conventional framework and one with
the scalar off-shell coupling.

In the lowest order in $v/c$, this term gives the following
contribution to the charge density
\begin{equation}
 \delta \rho_{s,con}
 \simeq \frac{\nu_s g_s^2}{2m}
 \left\{ \hat{e}_1,  \vec{\tau}_1 \cdot \vec{\tau}_2 \right\}\,
 D_s(p_2'- p_2)  \, ,
\label{delrhocons}
\end{equation}
where $\hat{e}_1$ is the charge of the first nucleon
as an operator in isospin space. This term has a structure which is very
similar to the retardation contribution from sigma exchange estimated in
Ref.~\cite{HAS}, which was found to give a nonnegligible
contribution to the deuteron and trinucleon form factors. We therefore
expect (\ref{delrhocons}) also to give a nonnegligible effect.

\section{Off-shell effects and nonrelativistic expansions}

It is instructive to compare our results for off-shell scalar exchange
to results which have been previously obtained from the study of
off-shell pion exchanges.  These studies have been
carried out in the framework of conventional perturbative expansions
of boson-nucleon vertices in powers of $v/c$, time-ordered diagrams for
potentials and currents, and wave functions obtained from
Schr\"odinger-like equations. In this framework, the nucleons are on
their mass shells, but energy is not conserved at the vertices and
therefore the potential and the current operators do not  commute with
the hamiltonian. The definition of these operators off-shell is not
unique, is subject to ambiguities, and varies for different
methods.  For pion exchange \cite{Fr,HG,GA} it has been shown
\cite{Ad,Fr} that all the results of various methods are, to leading
relativistic order, covered by a generic formula
\begin{equation}
 A(\tilde{\mu}) = i \tilde{\mu} \,  [ A_{n.r.}, U ]
  + \Delta A(\mu) \, ,
\label{unit}
\end{equation}
where $A_{n.r.}$ is a corresponding nonrelativistic operator,
$U$ is hermitian interaction-dependent operator $\sim (v/c)^2$,
and a parameter $\tilde{\mu}$ depends on the particular method used [see
below (\ref{mutil})] and the PS-PV mixing parameter
$\mu$, defined in (\ref{gammaps}) with $\nu_{ps} \to \mu$ for pions.  
The last term on the r.h.s.~of (\ref{unit}), $\Delta A(\mu)$, follows from
the explicit dependence of the {\it underlying lagrangian}  on the
mixing parameter $\mu$, at the one-pion exchange
level it is just the nonminimal contact
exchange current.  If the chiral lagrangian is employed
$\Delta A(\mu)= 0$. The commutator terms generate {\it
additional } exchange currents, a contribution to {\it
one-pion-exchange} potential, and, if two-pion exchanges are
included, also two-pion-exchange $NN$ and $3N$ potentials.

In this section we show, that for scalars there is no
additional unitary freedom and terms analogous to the commutator
in (\ref{unit}) do not appear. Therefore, the one-scalar-exchange
potential does not depend on the off-shell parameter $\nu$ and
the explicitly $\nu$-dependent potentials and currents derived
in this paper correspond to the last term in (\ref{unit}) and they
appear because the original lagrangian does depend on $\nu$.

Let us first recall the results for the pseudoscalar
mesons \cite{Fr,HG,GA}. From the lagrangian with mixture of the
PS and PV couplings one gets
\begin{equation}
 i \partial_t \psi = \biggl\{ \vec{\alpha} \cdot \vec{p} + \gamma^0  m
 + i g (1- \mu) \gamma^0  \gamma^5 \Phi - \frac{\mu g}{2m}
 (\vec{\sigma} \cdot \vec{\nabla} \Phi ) - \frac{\mu g}{2m}
 \gamma^5 (\partial_t \Phi ) \biggr\} \psi = H \psi \, .
\label{direqpi}
\end{equation}
Removing the odd operators by a standard FW procedure
(see e.g., chapter 4 of \cite{BD}), one gets
\begin{eqnarray}
 H^{int}_{FW}& =& - \frac{g}{2m} \biggl(
 (\vec{\sigma} \cdot \vec{\nabla} \Phi ) - \frac{1}{4m^2}
 \{ \vec{p}^{\, 2}, (\vec{\sigma} \cdot \vec{\nabla} \Phi ) \}
  + (1+\mu+c(1-\mu)) \frac{1}{4m}\gamma^0  \{ \vec{\sigma} \cdot \vec{p}
 \, , (\partial_t \Phi ) \} \nonumber\\
 &&+
 (\mu+ c(1-\mu)] \frac{i}{8m^2} \{ \vec{\sigma} \cdot \vec{p}\,  ,
 [\, \vec{p}^{\, 2} \, , \Phi ] \} \biggr) \, ,
\label{hintpi}
\end{eqnarray}
where $c$, defined as in \cite{HG,GA}, is so-called Barnhill
parameter \cite{Barn} which determines in which
order are the odd terms eliminated. In the momentum space, the
vertex function for $i$-th nucleon interacting with pion  derived
from (\ref{hintpi}) is
\begin{eqnarray}
 \Gamma(p'_i, p_i) &= & \frac{i}{2m}
  \biggl[ (\vec{\sigma}_i \cdot \vec{q}_i )
  \left( 1 + \frac{\vec{p}^{\, '\, 2}_i +  \vec{p}_i^{\, 2} }{4m^2}
  \right)    \nonumber\\
 && -
  \vec{\sigma}_i \cdot  (\vec{p}^{\, '}_i + \vec{p}_i\, ) \,
  \frac{1}{4m} \biggl( \Delta E_i - (1+\mu+c(1-\mu)) (\Delta E_i - q_{0,i} )
  \biggr) \biggr] \, ,
\label{vertfwpi}
\end{eqnarray}
where $\Delta E_i = E'_i -E_i$ is the energy difference for final and
initial on-mass-shell nucleon and $q_{0,i}$ is an energy of the absorbed
meson. For pions,
the unitary freedom appears because one can extend the vertex
off-energy- shell, i.e., set $\Delta E_i \neq q_{0,i}$. In fact, it has
been shown
\cite{Ad,AGA} that various ways to construct two-nucleon operators
effectively replace
\begin{equation}
 \Delta E_i - q_{0,i} = \beta (E'_1 + E'_2 - E_1- E_2) =
                         \beta (E' -E) \, ,
\label{offener}
\end{equation}
where $\beta$ is an arbitrary parameter, introduced in \cite{AGA},
and for most techniques in use  $\beta=1/2$.
The unitary parameter $\tilde{\mu}$ from (\ref{unit}) is given by
\begin{equation}
 1+\tilde{\mu} = 2 \beta \left[ \mu+1 +c(1-\mu)  \right]  \, ,
\label{mutil}
\end{equation}
and it combines dependence on the off-shell parameter
$\mu$ with dependence on $c$ and $\beta$, which  parametrize all of the
various  ways of doing FW transformations and constructing
interaction-dependent operators. For any value of $\tilde{\mu}$ one
gets a hermitian hamiltonian and a description conforming with
approximate Lorentz and gauge invariance.

Let us now consider the case of scalar mesons.
 From the lagrangian (\ref{lagrs0})
we obtain the equation for the time dependence of the nucleon
field
\begin{eqnarray}
 i \partial_t \psi &=& \biggl\{ \vec{\alpha} \cdot \vec{p} + \gamma^0 m
 - g (1- \nu) \gamma^0  \Phi +
  \frac{a}{2} \left\{ \vec{\alpha} \cdot \vec{p}, \, \Phi \right\}
 - \frac{a}{2} i (\partial_t \Phi ) - a \Phi i \partial_t \biggr\} \psi
\nonumber\\
& \simeq & \biggl\{ \vec{\alpha} \cdot \vec{p} +  \gamma^0 m - g  \gamma^0
\Phi +
  \frac{a}{2} \left[ \vec{\alpha} \cdot \vec{p}, \, \Phi \right]
  - \frac{a}{2} i (\partial_t \Phi ) \biggr\} \psi = H\, \psi \, ,
\label{direqs}
\end{eqnarray}
where the second form follows from approximation
\begin{equation}
i \partial_t \psi
\simeq  (\vec{\alpha} \cdot \vec{p} + \gamma^0  m ) \psi
\label{psikin}
\end{equation}
on r.h.s. (recall that only terms linear in $\Phi$ are retained).
Removing the odd operators, we obtain from the second form
\begin{equation}
 H^{int}_{FW} = - g  \gamma^0 \biggl[ \Phi -
 \frac{1}{8 m^2} \left\{ \vec{\sigma} \cdot \vec{p} \, ,
 \left\{ \vec{\sigma} \cdot \vec{p} \, , \Phi \right\} \right\} \biggr]
 + \frac{\nu g}{2 m} \biggl( - i (\partial_t \Phi) +  \gamma^0
 \left[ \frac{\vec{p}^{\, 2}}{2m}\, , \Phi\right] \biggr) \, .
\label{hfw1}
\end{equation}
Note that, unlike for pions,  there is no Barnhill
freedom at the order considered. The reason is that the
interaction-dependent odd terms of the untransformed hamiltonian
(\ref{direqs}) are of the order $\sim 1/m$ and hence the
corresponding unitary transformation would generate the
contributions $\sim 1/m^3$, while we keep
only  the terms up to  $\sim 1/m^2$.
In momentum space, the hamiltonian (\ref{hfw1}) generates
the $sNN$ vertex for i-th nucleon
\begin{equation}
 \Gamma(p'_i, p_i) = \biggl[ 1 - \frac{1}{8m^2} \left(
 (\vec{p}^{\, '}_i + \vec{p}_i\, )^2 + 2i \vec{\sigma}_i \cdot
  (\vec{p}^{\, '}_i  \times \vec{p}_i\, \right)
  - \frac{\nu}{2m} (\Delta E_i - q_{0,i} ) \biggr] \, .
\label{vertfws}
\end{equation}

For our scalar exchange this off-energy shell extension (\ref{offener})
is not allowed. The point is, that if the last $\nu$-dependent term
is present in (\ref{vertfws}), the vertex and the one-scalar exchange
potential derived from it are not hermitian, i.e.,
$\Gamma(p'_i, p_i) \neq \Gamma(p_i, p'_i)^*$. It is clear already
from (\ref{direqs}), since the hamiltonian defined in (\ref{direqs})
is hermitian only if
\begin{equation}
i (\partial_t \Phi) =  \gamma^0
 \left[ \frac{\vec{p}^{\, 2}}{2m}\, , \Phi\right] \, ,
\end{equation}
that is only if energy is conserved at the vertex.

We could attempt to re-write the eq. (\ref{direqs}) with the hamiltonian
which is equivalent to the old one on energy shell and which is
hermitian
\begin{equation}
 i \partial_t \psi = \biggl\{ \vec{\alpha} \cdot \vec{p} +  \gamma^0 m
 - g (1- \nu) \gamma^0  \Phi +
  \frac{a}{2} \left\{ \vec{\alpha} \cdot \vec{p}, \, \Phi \right\}
 - \frac{a}{2} \left\{ i \partial_t \, , \Phi \right\} \biggr\} \psi
 = H' \psi \, ,
\label{direqp}
\end{equation}
but with the help of (\ref{psikin}) one easily sees that
to first order in $\Phi$, the hamiltonian
$H'$ is equivalent to the standard $\nu$-independent one
\begin{equation}
 H' \simeq \vec{\alpha} \cdot \vec{p} + \gamma^0  m - g  \gamma^0  \Phi \, ,
\end{equation}
and it gives only the standard part of (\ref{vertfws}). This can also be
checked by straightforward FW transformation of (\ref{direqp}).
Therefore, we conclude that there is no consistent way to continue
the  $sNN$ vertex off-energy shell. Consistent operators are obtained only
if one puts $\beta=0$, i.e., $\tilde{\mu}=-1$ in (\ref{offener})
and (\ref{mutil}), respectively, as in the $S$-matrix approach
\cite{ATA}. Then, the nucleon potentials and exchange currents are
identified with the straightforward Taylor expansion in powers of
of $v/c \sim p/m$ of the corresponding
Feynman amplitudes. This identification is made in previous sections.
In particular, only the standard $\nu$-independent part
of the $sNN$ vertex contributes to one-scalar-exchange
potential and this potential  does not depend
on  $\nu$ even when the relativistic corrections
are included and the potential is considered off-energy-shell.

\section{Conclusions}

The lagrangian with the off-shell vertices can be conveniently
related to the more conventional one by means of the nucleon field
redefinition. The transformed lagrangian contains simpler $bNN$ vertices,
but has complicated contact multimeson terms.
One can also start from a conventional lagrangian and introduce
off-shell couplings and some contact terms. Notice that this way
one never gets the coupling with off-shell extension both before
and after meson is emitted.

The nonlinear interactions generate triangle and bubble diagrams
for the $NN$ interaction, complicated three-nucleon interactions
and meson exchange currents. We list leading order contributions
to these operators.

 For the scalar, pseudoscalar and isoscalar vector mesons one can
 completely transform
away the interaction terms with derivatives of the nucleon field, for the
isovector vector mesons this is possible only at the lowest order,
linear in the vector field.

Since in the low and intermediate energy region the importance of
nucleon operators is increasing with decreasing exchanged meson mass,
it would be interesting to numerically estimate the effect of these
lowest order operators, since they might represent a large
part of the difference between a model with off-shell coupling and
one with conventional vertices.

Unlike for the pion case, for scalars there is
no additional unitary freedom dependent on the off-shell parameter
$\nu$ in the framework of
a hamiltonian formalism with a $v/c$ expansion.

\acknowledgments

We would like to acknowledge the support of the DOE through
Jefferson Laboratory and one of us (F.G.) also gratefully acknowledges
support from DOE grant No. DE-FG02-97ER41032. J.A. and J.W.V.O. thank
Institute for Nuclear Theory  at the University of Washington for its
hospitability during our stay while this work was completed. We also thank
S.A.~Coon for reading the manuscript and making helpful remarks.

This manuscript is sent to {\it Few Body Systems\/} in celebration of
Professor John Tjon's 60th birthday.  We have all learned much from
Prof.~Tjon and want to use this opportunity to thank him for sharing his
ideas and physical insight with us.

\renewcommand{\theequation}{A.\arabic{equation}}
\setcounter{equation}{0}
\section*{Appendix A: Pseudoscalar and vector exchanges}

In this appendix we demonstrate how the removal of the off-shell
\cite{StG} coupling works for pseudoscalar and vector mesons.
In a generic form the lagrangian with an off-shell coupling reads
\begin{eqnarray}
  {\cal L} & = & {\cal L}^{kin}_B + {\cal L}_{\gamma bb} +
  {\cal L}^{kin}_N (\psi ) + {\cal L}_{bNN} (\psi )
  + {\cal L}_{\gamma NN} (\psi ) + {\cal L}_{\gamma NNb}(\psi )
  \, , \label{lagr0}\\
  {\cal L}_{bNN} (\psi ) &=&   g \bar{\psi} \Gamma \Phi \psi
  +  \frac{a}{2}  \bar{\psi} (\Gamma \Phi)
  ( i \gamma_{\mu} \partial^{\mu} \psi - m \psi  ) -
  \frac{a}{2}
  ( i (\partial^{\mu} \bar{\psi}) \gamma_{\mu} + m \bar{\psi}  )
  (\Gamma \Phi) \psi       \, ,
  \label{lbnn}\\
   {\cal L}_{\gamma NNb} (\psi )  &=&   \frac{a}{2} \bar{\psi}
   \left\{  \Lambda_0^{\mu} \, ,
    \Gamma \Phi \right\} \psi \, A_{\mu} \, \, ,
\label{lgnnb}
\end{eqnarray}
where we  use $b= s, v, ps$ to denote different mesons and
$B= b, \gamma$ stands for both mesons and photons.
The $bNN$ vertex $\Gamma$\ contains some Dirac matrix structure,
and it can carry a Lorentz index for spin 1 mesons and an isospin
index for isospin 1 mesons. In more detail, we take
for scalar mesons $\Gamma_s = 1$, for pseudoscalar mesons
$\Gamma_{ps}= - i \gamma^5$ and for vector mesons $\Gamma_v^{\mu} =
\gamma^{\mu} $. The $vBB$ vertex could also contain an anomalous tensor
coupling, but the off-shell extension of this part has not been considered
in \cite{StG} and we omit the anomalous
coupling  here for the sake of simplicity .
The  $bNN$ vertex function in momentum space reads
\begin{equation}
 \tilde{\Gamma} (p', p) = \Gamma(p', p) +
 \frac{\nu}{2 m} \left( ( \hat{p}' -m) \Gamma(p', p)+
 \Gamma(p', p) ( \hat{p} -m) \right) \, ,
\label{offp}
\end{equation}
For the pseudoscalar mesons the off-shell couplings (\ref{offp}) is just
the usual pseudovector vertex,
\begin{equation}
 \tilde{\Gamma}_{ps}(p', p) = - i\, \Biggl( (1- \nu_{ps})
 \gamma^5 + \frac{\nu_{ps}}{2 m} (\hat{p}' - \hat{p})  \gamma^5 \Biggr)
  \, ,
\label{gammaps}
\end{equation}
 which also follows from adding the
total derivative to (\ref{lagr0}). For pions $\nu_{ps}$ is usually 
named  $\mu$. This is possible since
$\{ \gamma^\mu, \gamma^5 \}= 0$ and it completely removes the
derivatives of the nucleon fields from the interaction terms,
introducing instead terms with derivatives of the meson field,
i.e., replacing the difference of nucleon momenta by the meson
momentum $q= p' -p$.
Hence, the field redefinition removing this part of vertex should
be equivalent to the usual chiral rotation which dials between PS and PV
couplings. Nevertheless, let us recover these results proceeding in
a same way as used for scalar and vector mesons, though for pseudoscalar
mesons the procedure appears somewhat artificial.

To transform away the off-shell $bNN$, we
redefine the nucleon field with the help of the
function $F= F(\Gamma \Phi)$, obeying
\begin{equation}
\gamma^0 F^*(\Gamma \Phi) \gamma^0 = F (\Gamma \Phi) \, ,
\label{frel1}
\end{equation}
so that
\begin{eqnarray}
 \psi(x) &=& F(\Gamma \Phi (x))\, \psi'(x) \, , \\
 \bar{\psi}(x) &=&  \bar{\psi'}(x) F(\Gamma \Phi (x))\, \, .
\label{psi3}
\end{eqnarray}
In terms of $\psi'$
\begin{eqnarray}
{\cal L} &=&
 \frac{i}{2} \bar{\psi'} F ( 1+ a \Gamma \Phi) \gamma^{\mu} F
  ( \partial_{\mu} \psi' ) -
\frac{i}{2} (\partial_{\mu} \bar{\psi'}) F \gamma^{\mu}
 ( 1+ a \Gamma \Phi) F \psi'
 - m \bar{\psi'} F ( 1+ a \Gamma \Phi) F \psi'
   \nonumber\\
&&  + g  \bar{\psi'} F^2 (\Gamma \Phi) \psi'
   + \frac{i}{2} \bar{\psi'} F ( 1+ a \Gamma \Phi) \gamma^{\mu}
  (\partial_{\mu} F ) \psi' - \frac{i}{2}   \bar{\psi'}
 (\partial_{\mu} F ) \gamma^{\mu}  F ( 1+ a \Gamma \Phi) \psi'
\nonumber\\
&& + \bar{\psi}' F \, \Lambda^{\mu} \, F \psi' \, A_{\mu}
   + \frac{a}{2} \bar{\psi}'  F \,
   \left\{ \Lambda_0^{\mu} \, , \Gamma \Phi \right\} \,
     F \psi' \, A_{\mu} \, .
\label{ltran}
\end{eqnarray}
Requiring that the first line in (\ref{ltran}) equals the nucleon kinetic
lagrangian ${\cal L}^{kin}_N (\psi')$ implies
\begin{equation}
 \gamma^\mu = F ( 1+ a \Gamma \Phi) \gamma^\mu F =
              F \gamma^\mu ( 1+ a \Gamma \Phi) F \, ,
\label{vertf}
\end{equation}
which can be satisfied only if $ [ \gamma^\mu \, , \Gamma ] = 0$.
However, this does not mean that it is possible to transform
away interactions with derivatives of nucleon fields only in this
case: the offending terms might be eliminated by adding a total
derivative. Indeed, let us denote
\begin{eqnarray}
 \Gamma^\mu_1 &=&  F ( 1+ a \Gamma \Phi) \gamma^{\mu} F \, , \\
 \Gamma^\mu_2 &=&  F \gamma^{\mu} ( 1+ a \Gamma \Phi) F \, ,
\end{eqnarray}
and add to the lagrangian (\ref{ltran}) a total derivative
\begin{equation}
  \frac{i}{2} \partial_\mu \biggr[ \bar{\psi'}
   \left( f  \Gamma^\mu_1 - \Gamma^\mu_2 f \right) \psi' \biggl] \, ,
\label{deriv}
\end{equation}
where $f = f(\Gamma \Phi )$ is to be determined to allow reduction of
all terms with derivatives of nucleon fields to ${\cal L}^{kin}_N (\psi')$.
These terms now equal
\begin{equation}
 \frac{i}{2} \bar{\psi'} \left( (f+1) \Gamma^\mu_1- \Gamma^\mu_2 f \right)
  ( \partial_{\mu} \psi' ) - \frac{i}{2} (\partial_{\mu} \bar{\psi'})
 \left( \Gamma^\mu_2 (f+1) - \Gamma^\mu_1  f \right) \psi' \, .
\end{equation}
Requiring that
\begin{equation}
 \gamma^\mu =   (f+1) \Gamma^\mu_1- \Gamma^\mu_2 f
            =   \Gamma^\mu_2 (f+1) - \Gamma^\mu_1  f   \, ,
\label{kingam}
\end{equation}
leads to a condition
\begin{equation}
    (2f+1) \Gamma^\mu_1 = \Gamma^\mu_2 (2f+1)  \, ,
\end{equation}
or, assuming existence of $F^{-1}$, to the relations
\begin{equation}
    (2f+1) ( 1+ a \Gamma \Phi) \gamma^\mu =
     \gamma^\mu ( 1+ a \Gamma \Phi)  (2f+1)  \, ,
\end{equation}
which is satisfied with
\begin{equation}
   f (\Gamma \Phi) =
 \frac{c -a \Gamma \Phi}{2( 1+ a \Gamma \Phi)} \, ,
\label{solf}
\end{equation}
where $c$ is an arbitrary number, which can be set to $c=0$.
Using the solution (\ref{solf}) in the constraint (\ref{kingam})
leads to an equation for $F$
\begin{equation}
  \gamma^\mu = F \biggl( \gamma^\mu + \frac{a}{2}
  \left\{\gamma^\mu\, , \Gamma \Phi \right\} \biggr) F  \, .
\label{frel2}
\end{equation}
Assuming that the solution $F$ of (\ref{frel2}) exists,
the transformed lagrangian (\ref{ltran})  with the total
derivative (\ref{deriv}) added  and with the function
$f(\Gamma \Phi)$ given in (\ref{solf}) is
\begin{eqnarray}
{\cal L} &=&
 \frac{i}{2} \bar{\psi'} \gamma^{\mu}( \partial_{\mu} \psi' ) -
 \frac{i}{2} (\partial_{\mu} \bar{\psi'}) \gamma^{\mu} \psi'
 - m \bar{\psi'} F ( 1+ a \Gamma \Phi) F \psi'
   + g  \bar{\psi'} F^2 (\Gamma \Phi) \psi' \nonumber\\
&& + \frac{i}{2} \bar{\psi'} \Biggl(
  \gamma^{\mu} F^{-1} (\partial_{\mu} F ) -
  (\partial_{\mu} F ) F^{-1} \gamma^{\mu} +
 \frac{a}{2} F \, [\gamma^{\mu}\, , (\Gamma \partial_{\mu} \Phi ) ]\, F
  \Biggr) \psi' \nonumber\\
&& + \bar{\psi}' F \, \Lambda^{\mu} \, F \psi' \, A_{\mu}
   + \frac{a}{2} \bar{\psi}'  F \,
   \left\{ \Lambda_0^{\mu} \, , \Gamma \Phi \right\} \,
     F \psi' \, A_{\mu} \, .
\label{ltran1}
\end{eqnarray}

For scalar mesons $\Gamma \to \Gamma_s=1$ and hence
$\Gamma$ commutes with $\gamma^\mu$.  The matrix $\gamma^\mu$
can be factorized from (\ref{frel2}), we get
$F_s^2 = 1+ a_s \Phi_s $, and from (\ref{ltran1}) one
immediately recovers  (\ref{lagrs}).

For pseudoscalar mesons $\Gamma \to \Gamma_{ps}= - i \gamma^5$,
and $\Gamma$ anticommutes with $\gamma^\mu$.
Considering now
$F \to F_{ps}(- i \gamma^5 \Phi_{ps})$, for which  $F_{ps} \gamma^\mu =
\gamma^\mu F_{ps}^*$, we obtain from (\ref{frel2}) the
relation
\begin{equation}
F_{ps} F_{ps}^* =1 \, ,
\label{frel3}
\end{equation}
i.e., $F_{ps}^{-1} =  F_{ps}^*$.
Using these properties of $F_{ps}$ we obtain from (\ref{ltran1})
\begin{eqnarray}
{\cal L} &=&
{\cal L}_{kin}^N (\psi' )
 + \frac{\nu_{ps} g_{ps}}{2 m}
 \bar{\psi'} \gamma^{\mu} \gamma^5
  F^* (\partial_{\mu} \Phi_{ps}) F  \psi'
 - i g_{ps} (1- \nu_{ps}) \bar{\psi'} F_{ps}^2 \gamma^5 \Phi_{ps}\, \psi'
\nonumber\\
&& + m  \bar{\psi'} (1- F_{ps}^2 )  \psi' +
 \frac{i}{2} \bar{\psi'} \biggl(
 F_{ps} \gamma^{\mu}  (\partial_{\mu}  F_{ps}) -
 (\partial_{\mu}  F_{ps}) \gamma^{\mu}  F_{ps}
  \biggr) \psi' \nonumber\\
&&+ \bar{\psi'} F_{ps} \Lambda^\mu F_{ps}\, \psi' A_\mu -
 i \frac{a_{ps}}{2} \bar{\psi'} F_{ps}
 \left\{ \Lambda_0^\mu, \gamma^5 \Phi_{ps} \right\} F_{ps}\, \psi' A_\mu
  \, .
\label{lpscal}
\end{eqnarray}
The most general form  of $F_{ps}(- i \gamma^5 \Phi_{ps})$, satisfying
the constraints (\ref{frel1}), and (\ref{frel3}) is
\begin{equation}
F_{ps}(- i \gamma^5 \Phi_{ps}) =
 \exp \left( i \eta \frac{g_{ps}}{2m} \Phi_{ps} \gamma^5
 \, f(\Phi_{ps}^2) \right)
\, ,
\label{fpscal}
\end{equation}
where $\eta$ is real number and $ f^*(\Phi_{ps}^2)= f(\Phi_{ps}^2)$,
normalized to $f(0)= 1$.
Up to the quadratic terms in $\Phi_{ps}$ the lagrangian
(\ref{lpscal}) does not depend on the form of $f(\Phi_{ps}^2)$
and we can replace $f(\Phi_{ps}^2) \to f(0)= 1$:
\begin{eqnarray}
{\cal L}  &\simeq&
{\cal L}_{kin}^N (\psi' )
+ \frac{(\nu_{ps}-\eta) g_{ps}}{2 m}
 \bar{\psi'} \gamma^{\mu} \gamma^5 (\partial_{\mu} \Phi_{ps})  \psi'
- i g_{ps} (1- \nu_{ps}+\eta) \bar{\psi'} \gamma^5 \Phi_{ps} \psi'
\nonumber\\
&& + \eta ( 1 - \nu_{ps} + \frac{\eta}{2} )
 \frac{g_{ps}^2}{m}\, \bar{\psi'} \Phi_{ps}^2 \psi'
 + \eta (2 \nu_{ps} - \eta ) \frac{g_{ps}^2}{4 m^2}\,
\bar{\psi'}\gamma^{\mu} \vec{\tau} \cdot
\vec{\Phi}_{ps} \times (\partial_{\mu}  \vec{\Phi}_{ps})\, \psi' \nonumber\\
&& + \bar{\psi'} \Lambda^\mu \psi' \, A_\mu
+ i \frac{g_{ps}}{2m}\, \bar{\psi'} \left\{
 (\eta- \nu_{ps}) \Lambda_0^\mu + \eta \Delta \Lambda^\mu \, ,
 \gamma^5 \Phi_{ps} \right\} \psi' \, A_\mu
  \, .
\label{lpscala}
\end{eqnarray}
The transformed lagrangian contains a $psNN$ vertex with the PS-PV mixing
(the mixing parameter is $\nu_{ps}-\eta$) and the quadratic contact
vertices of the standard form. The e.m.\ coupling with $\Lambda_0^\mu$
is just the the usual Kroll-Rudermann coupling \cite{KR}, it is
obtained by the minimal coupling from the $psNN$ vertex with a derivative
and it contains the same  factor $(\nu_{ps}- \eta)$.
 For pseudoscalar mesons the
interactions with derivatives of nucleon fields is replaced by
the PV form of meson-nucleon coupling with derivative of the meson
field. Thus, in this case transforming away the off-shell coupling
means eliminating the PV $psNN$ vertex, which is achieved if
one sets $\eta = \nu_{ps}$. The transformed lagrangian then simplifies
to
\begin{eqnarray}
{\cal L}  &\simeq&
{\cal L}_{kin}^N (\psi' )
- i g_{ps} \bar{\psi'} \gamma^5 \Phi_{ps} \psi'
\nonumber\\
&& + \nu_{ps} ( 1 - \frac{\nu_{ps}}{2} )
 \frac{g_{ps}^2}{m}\, \bar{\psi'} \Phi_{ps}^2 \psi'
 + \frac{g_{ps}^2 \nu_{ps}^2}{4 m^2}\, \bar{\psi'}\gamma^{\mu}
 \vec{\tau} \cdot \vec{\Phi}_{ps} \times (\partial_{\mu}  \vec{\Phi}_{ps})\,
  \psi' \nonumber\\
&&  + \bar{\psi'} \Lambda^\mu \psi' \, A_\mu
 + i \frac{ \nu_{ps} g_{ps}}{2m}\, \bar{\psi'}
 \left\{ \Delta \Lambda^\mu \, ,
 \gamma^5 \Phi_{ps} \right\} \psi' \, A_\mu   \, .
\label{lpscala1}
\end{eqnarray}

For the vector mesons $\Phi \to v_\mu$, $\Gamma \to \gamma^\mu$ and the
off-shell vertex in momentum space reads
\begin{eqnarray}
 \tilde{\Gamma}^{\mu}_{v}(p', p) &= &
 \gamma^\mu + \frac{\nu_v}{2 m}
 \left( (\hat{p}'- m)\gamma^\mu + \gamma^\mu (\hat{p} -m ) \right)
 \nonumber\\
  &= &  (1-\nu_v) \gamma^\mu + \frac{\nu_v}{2 m}
   \left( (p' + p)^\mu + i \sigma^{\mu \rho} (p' -p)_\rho  \right)
  \, .
\end{eqnarray}
From (\ref{frel2}) we get for the vector mesons
\begin{equation}
  \gamma^\mu = F_v  ( \gamma^\mu +  a_v v^\mu ) F  \, .
\label{frelv}
\end{equation}

Let us first consider isoscalar vector mesons. In this case,
$v_\mu$ commutes with $\hat{v}$, and hence with $F_v(\hat{v})$. Therefore,
we can multiply (\ref{frelv}) by $v_\mu$ and find the solution
\begin{equation}
 F_v (\hat{v})= \frac{1}{\sqrt{1 + a_v \hat{v}}} \, ,
\end{equation}
which as a matrix in the Dirac space is defined by its Taylor series.
The transformed lagrangian reads
\begin{eqnarray}
{\cal L} &=& {\cal L}_N^{kin}(\psi')
    + g_v  \bar{\psi'} \frac{\hat{v}}{1 + a_v \hat{v}}  \psi' \nonumber\\
&& + \frac{i}{2} \bar{\psi'} \Biggl(
  \gamma^{\mu} F_v^{-1} (\partial_{\mu} F_v ) -
  (\partial_{\mu} F_v ) F_v^{-1} \gamma^{\mu} +
 \frac{a_v}{2} F_v \, [\gamma^{\mu}\, , ( \partial_{\mu}\hat{v}) ]\, F_v
  \Biggr) \psi' \nonumber\\
&& + \bar{\psi}' F_v \, \Lambda^{\mu} \, F_v \psi' \, A_{\mu}
   + \frac{a_v}{2} \bar{\psi}'  F_v \,
   \left\{ \Lambda_0^{\mu} \, , \hat{v} \right\} \,
     F_v \psi' \, A_{\mu} \nonumber\\
 & \simeq & {\cal L}_N^{kin}(\psi')+ g_v \bar{\psi'} \hat{v}  \psi'
 -  \frac{a_v}{2} \bar{\psi}'  \left\{ \Delta \Lambda^{\mu} \, ,
 \hat{v} \right\} \, \psi' \, A_{\mu} \nonumber\\
&& - \frac{\nu_v g_v^2}{m} \bar{\psi'} (v_\rho v^\rho) \psi'
 - i \frac{\nu_v^2 g_v^2}{8m^2}  \bar{\psi'}
 \biggl( \hat{v} \gamma^{\mu} (\partial_\mu  \hat{v} ) -
 (\partial_\mu  \hat{v} ) \gamma^{\mu}  \hat{v}  \biggr) \psi' \, .
\label{ltranv1}
\end{eqnarray}

For isovector vector mesons a solution to all order in meson
fields does not seem to exist. The problem is that in this case
$v_\alpha = \vec{\tau} \cdot \vec{v}_\alpha$ and the components
of the vector field do not commute
\begin{equation}
[v_\alpha , v_\beta ] = 2 i \, \vec{\tau} \cdot
 ( \vec{v}_\alpha \times \vec{v}_\beta ) \, .
\end{equation}
Still, if only terms linear in the meson field are retained,
the choice $F_v \simeq 1 - a_v \hat{v}/2 $ eliminates the interaction
terms with derivatives of nucleon fields. Up to a quadratic order,
the most general form of $F_v$ is
\begin{equation}
F_v \simeq 1 - \frac{a_v}{2} \hat{v} + c a_v^2 \hat{v} \hat{v} +
          d a_v^2 v_\alpha \hat{v} \gamma^\alpha \, ,
\end{equation}
and it is easy to check that it does not solve (\ref{frelv}) up to
the quadratic order for any choice of $c$ and $d$. This means that
any field redefinition leaves some interaction terms quadratic in
vector field and containing the derivatives of nucleon field.

To sum it up, the linear approximation to function $F$ is in all
cases (with $\eta = \nu_{ps}$ for pseudoscalar mesons) given by
\begin{equation}
 F \simeq 1 - \frac{a}{2} \Gamma \Phi \, ,
\end{equation}
and the transformed lagrangian up to this order reads
\begin{equation}
  {\cal L} (\psi')  = {\cal L}_N^{kin}(\psi') +
  g  \bar{\psi'} \Gamma \Phi  \psi'
  -  \frac{a}{2} \bar{\psi}'  \left\{ \Delta \Lambda^{\mu} \, ,
  \Gamma \Phi  \right\} \, \psi' \, A_{\mu}  \, ..
\end{equation}
For scalar, pseudoscalar and vector isoscalar mesons the transformation
can be carried out to all orders and different contact interactions appear.
For vector isovector mesons it seems impossible to transform away
the interaction terms with derivatives of nucleon fields to higher
than linear order in meson field. Although the closed solution exists
for most types of mesons, if more than one off-shell vertex is considered
one has to resort to approximate solution approximating $F$ by a power series.

\end{document}